# Fokker-Planck Theory of Nonequilibrium Systems Governed by Hierarchical Dynamics

Sumiyoshi Abe

Department of Physical Engineering, Mie University, Mie 514-8507, Japan

**Abstract** Dynamics of complex systems is often hierarchically organized on different time scales. To understand the physics of such hierarchy, here Brownian motion of a particle moving through a fluctuating medium with slowly varying temperature is studied as an analytically tractable example, and a kinetic theory is formulated for describing the states of the particle. What is peculiar here is that the (inverse) temperature is treated as a dynamical variable. Dynamical hierarchy is introduced in conformity with the adiabatic scheme. Then, a new analytical method is developed to show how the Fokker-Planck equation admits as a stationary solution the Maxwellian distribution modulated by the temperature fluctuations, the distribution of which turns out to be determined by the drift term. A careful comment is also made on so-called superstatistics.

**Keywords**   Hierarchical dynamics · Fluctuating dynamical temperature · Fokker-Planck equation

# 1 Introduction

Nonequilibrium complex systems are often governed by hierarchically organized dynamics composed of different subdynamics on different time scales. Decomposition of the total dynamics into subdynamics may be relatively simple if the time scales characterizing them are largely separated. Consider two random variables, say $X$ and $Y$, and suppose the dynamics of $X$ to be much faster than that of $Y$. The joint probability distribution $P(x,y)$ with $x$ and $y$ respectively being the realizations of $X$ and $Y$ is expressed by the Bayes rule as follows: $P(x,y) = P(x|y)P(y)$, where $P(x|y)$ is the conditional probability distribution of $X$ to have the value $x$, given a value of $Y$ to be $y$, and $P(y)$ is the marginal distribution of $Y$.

A corresponding physical situation in nonequilibrium statistical mechanics is as follows. A system is divided into small spatial cells, each of which still has to contain a sufficiently large number of particles, so that it is in a local equilibrium state characterized by inverse temperature, $\beta \equiv 1/(k_B T)$, where $k_B$ is the Boltzmann constant. There is the temperature gradient between two neighboring cells, which creates the heat flux. Such a flux may not bring significant effects if a system is simple. In a complex system, however, elements contained are strongly correlated over a great distance, and the flux is not ignorable, in general. So, the fluctuations of $\beta$ in each cell is supposed to be nonnegligible. Thus, $\beta$ should be regarded as the realization of the random variable, $B$.

The situation of physical interest is that relaxation of a cell from its arbitrary initial state to a local equilibrium state at a given value of $\beta$ is fast, whereas $\beta$ slowly varies according to a certain distribution, $f(\beta)$. The joint distribution describing a stationary state of the cell is then written as

$$P(\varepsilon_n, \beta) = P(\varepsilon_n | \beta) f(\beta), \qquad (1)$$

$$P(\varepsilon_n | \beta) = \frac{1}{Z(\beta)} \exp(-\beta \varepsilon_n), \qquad Z(\beta) = \sum_n \exp(-\beta \varepsilon_n). \qquad (2)$$

Here, $P(\varepsilon_n | \beta)$ describes a local equilibrium state of the cell, i.e., the conditional probability of finding the cell in the state with the energy, $\varepsilon_n$ (that is the $n$th realization of the random variable of the cell energy, $E$), given $\beta$.

The system mentioned above is governed by the dynamics with two hierarchies: that is, fast dynamics of $E$ and slow dynamics of $B$. The time scales characterizing these dynamics are largely separated. The randomness of $B$ is quenched during the relaxation of the state of the cell.

We note that, unlike $P(\varepsilon_n | \beta)$, $f(\beta)$ is introduced in Eq. (1) in an *ad hoc* manner. Without a physical principle for determining it, the discussion remains as formal as the Bayes rule.

In the present paper, we formulate a consistent approach, by which $f(\beta)$ is endowed with a mechanical implication. We consider this problem by employing an analytically tractable system: a Brownian particle moving through a background medium with slowly varying (inverse) temperature. This system does not seem to have been fully studied in the literature. A point to be emphasized is that here both the velocity of the particle and the (inverse) temperature are treated as random dynamical variables and satisfy coupled stochastic differential equations. Hierarchical dynamics is realized, with the help of the adiabatic scheme, for the Fokker-Planck equation associated with the stochastic differential equations. Then, developing a new method for obtaining a stationary solution of the multivariate Fokker-Planck equation [1,2], we explicitly show how $f(\beta)$ can be determined from the measurable drift term. We also make a comment on so-called superstatistics [3].

## 2  Kinetic Theory with Fluctuating Dynamical Temperature

Let us start our discussion with considering a two-dimensional vector-valued random variable

$$\mathbf{X} = \begin{pmatrix} X_1 \\ X_2 \end{pmatrix} \equiv \begin{pmatrix} V \\ \ln(B/b_0) \end{pmatrix}, \qquad (3)$$

where, $V$ is the velocity of a Brownian particle with unit mass moving through a fluctuating medium with slowly varying inverse temperature, $B$. $b_0$ is a positive constant having the dimension of $B$ and is henceforth set equal to unity. We take the logarithm of $B$, which is a nonnegative random variable. The dynamics of $\mathbf{X}$ is governed by a Langevin equation in the form of the following stochastic differential equation:

$$d\mathbf{X} = \mathbf{K}(\mathbf{X})\,dt + G\,d\mathbf{W}. \qquad (4)$$

Here, $\mathbf{K}(\mathbf{X})$ is a two-dimensional drift vector and $G$ is a $2\times 2$ matrix. $G$ might depend on **X**, in general. For the sake of simplicity, however, here we assume the process to be additive, so that $G$ is constant. We do not impose the fluctuation-dissipation relation since we are concerned with nonequilbrium states. A novel viewpoint here is that not only the velocity but also the inverse temperature is treated as a dynamical variable. One could also introduce spatial dependence of **X**, but we consider only the time dependence (i.e., assumption of spatial homogeneity). $d\mathbf{W}$ in Eq. (4) is the Wiener noise vector satisfying the Ito rule

$$dW_i\, dW_j = \delta_{ij}\, dt \qquad (i, j = 1, 2). \tag{5}$$

The correlation in the noise term in Eq. (4) is realized by the nondiagonal matrix, $G$.

In this system, there are four time scales. Let $\tau_V$ and $\tau_B$ be the time scales characterizing the dynamics of $V$ and $B$, while $\tau_V^*$ and $\tau_B^*$ be those of the noise terms appearing in the equations of $V$ and $B$, respectively. These are supposed to satisfy the following condition:

$$\tau_B \gg \tau_V \gg \tau_B^* > \tau_V^*. \tag{6}$$

$V$ is a fast variable, whereas $B$ is a slow one. Also, in order for the last inequality to be valid, the dominant element of $G$ appearing in the equation for $B$ has to be smaller than that in the equation for $V$. It is noted that $X_2$ may, in fact, be very slow since it is logarithmically related to the slow variable, $B$.

We assume that the equation for $B$ does not contain $V$. That is,

$$K_1 = K_1(X_1, X_2), \qquad K_2 = K_2(X_2). \tag{7}$$

This assumption implies that back reaction of the particle on the medium is negligible. This point is analogous to the Born-Oppenheimer approximation in quantum mechanics [4]. Therefore, the slow dynamics does not contain the fast variable, and the dynamical hierarchy is induced in this way.

Let $x_1 = v$ and $x_2 = \ln \beta$ be the realizations of $X_1 = V$ and $X_2 = \ln B$, respectively. The probability, $\hat{P}(\mathbf{x}, t) d^2 \mathbf{x}$, of finding the velocity and the logarithm of inverse temperature in the interval between $\mathbf{x} = (x_1, x_2)$ and $\mathbf{x} + d\mathbf{x} = (x_1 + dx_1, x_2 + dx_2)$ at time $t$ satisfies the Fokker-Planck equation [1,2]:

$$\frac{\partial \hat{P}(\mathbf{x}, t)}{\partial t} = -\sum_{i=1}^{2} \frac{\partial J_i(\mathbf{x}, t)}{\partial x_i}, \tag{8}$$

where $J_i(\mathbf{x}, t)$ is the probability current given by

$$J_i(\mathbf{x}, t) = K_i(\mathbf{x}) \hat{P}(\mathbf{x}, t) - \frac{1}{2} \sum_{j=1}^{2} D_{ij} \frac{\partial \hat{P}(\mathbf{x}, t)}{\partial x_j}, \tag{9}$$

and

$$D = G G^{\mathrm{T}} \tag{10}$$

is a constant symmetric diffusion matrix.

We are interested in a stationary solution of Eq. (8). The stationarity condition yields

$$\sum_{i=1}^{2} \frac{\partial J_i(\mathbf{x})}{\partial x_i} = 0. \tag{11}$$

To solve this equation, it is traditional to discuss boundary conditions (absorbing, reflecting, or periodic) on the probability current as well as the normalizability of the probability distribution. However, our purpose is to see if the solution of the form

$$\beta^{-1} \hat{P}(v, \ln \beta) \equiv P(v, \beta) = P(v|\beta) f(\beta), \tag{12}$$

$$P(v|\beta) = \frac{1}{Z(\beta)} \exp(-\beta \varepsilon), \qquad \varepsilon = \frac{1}{2} v^2, \tag{13}$$

with $\beta^{-1}$ in Eq. (12) being the Jacobian factor and $Z(\beta) = \sqrt{2\pi/\beta}$, can be obtained as a stationary solution of Eq. (11), which is the one-dimensional Maxwellian distribution (for unit particle number density), $P(v|\beta)$, modulated by the distribution of temperature fluctuations, $f(\beta)$. For this purpose, here we present the following new method. A basic idea is to set

$$J_i(\mathbf{x}) = \sum_{j=1}^{2} \frac{\partial h_{ij}(\mathbf{x})}{\partial x_j}, \tag{14}$$

where $h_{ij}(\mathbf{x})$ is an antisymmetric tensor. In the present case, $h_{ij}(\mathbf{x})$ is proportional to the two-dimensional Levi-Civita symbol ($\varepsilon_{ij} = -\varepsilon_{ji}$, $\varepsilon_{12} = 1$):

$$h_{ij}(\mathbf{x}) = \breve{h}(\mathbf{x}) \varepsilon_{ij}. \tag{15}$$

With this choice, Eq. (11) is automatically fulfilled. (A generalization of this method to a higher-dimensional case is straightforward.) Without losing generality, we can write

$$\breve{h}(\mathbf{x}) = \hat{h}(\mathbf{x})\hat{P}(\mathbf{x}).  \qquad (16)$$

Then, Eq. (14) together with Eqs. (9), (15), and (16) leads to the following coupled equations:

$$\left(K_1 - \frac{\partial \hat{h}}{\partial x_2}\right)\hat{P} - \frac{D_{11}}{2}\frac{\partial \hat{P}}{\partial x_1} - \left(\frac{D_{12}}{2} + \hat{h}\right)\frac{\partial \hat{P}}{\partial x_2} = 0, \qquad (17)$$

$$\left(K_2 + \frac{\partial \hat{h}}{\partial x_1}\right)\hat{P} - \left(\frac{D_{21}}{2} - \hat{h}\right)\frac{\partial \hat{P}}{\partial x_1} - \frac{D_{22}}{2}\frac{\partial \hat{P}}{\partial x_2} = 0. \qquad (18)$$

Substituting Eq. (12) into Eqs. (17) and (18), we obtain

$$\left(\frac{D_{12}}{2} + h\right)\beta\frac{df}{d\beta} = \left[K_1 - \beta\frac{\partial h}{\partial \beta} - \frac{3}{2}h - \frac{3D_{12}}{4} + \frac{D_{11}}{2}\beta v + \left(\frac{D_{12}}{4} + \frac{1}{2}\right)\beta v^2\right]f, \qquad (19)$$

$$\frac{D_{22}}{2}\beta\frac{df}{d\beta} = \left[K_2 + \frac{\partial h}{\partial v} - \frac{3D_{22}}{4} + \left(\frac{D_{12}}{2} - h\right)\beta v + \frac{D_{22}}{4}\beta v^2\right]f, \qquad (20)$$

where $h(v, \beta) \equiv \hat{h}(v, \ln\beta)$ and the symmetry, $D_{12} = D_{21}$ has been used. Since both of these are differential equations for $f(\beta)$, their consistency leads to the fact that there must exist a quantity, $A(\beta)$, satisfying the following equation:

$$\beta \frac{df(\beta)}{d\beta} = A(\beta) f(\beta). \tag{21}$$

This, in turn, yields

$$K_1 - \beta \frac{\partial h}{\partial \beta} - \frac{3}{2}h - \frac{3D_{12}}{4} + \frac{D_{11}}{2}\beta v + \left(\frac{D_{12}}{4} + \frac{1}{2}h\right)\beta v^2 = \left(\frac{D_{12}}{2} + h\right)A \tag{22}$$

$$K_2 + \frac{\partial h}{\partial v} - \frac{3D_{22}}{4} + \left(\frac{D_{12}}{2} - h\right)\beta v + \frac{D_{22}}{4}\beta v^2 = \frac{D_{22}}{2} A. \tag{23}$$

Now, differentiating Eq. (23) with respect to $v$ [and noting that $K_2 = K_2(\ln \beta)$], we have

$$\frac{\partial}{\partial v}\left(\frac{\partial h}{\partial v} - h\beta v\right) + \frac{D_{12}}{2}\beta + \frac{D_{22}}{2}\beta v = 0. \tag{24}$$

This equation admits as a solution the following linear function of $v$:

$$h(v, \beta) = \frac{D_{12}}{2} + \frac{D_{22}}{4} v, \tag{25}$$

which is actually independent of $\beta$. Therefore, from Eqs. (22) and (23), the elements of the drift vector are found to be given as follows:

$$K_1(v, \ln \beta)$$
$$= D_{12}\left(\frac{3}{2} + A(\beta)\right) + \left(\frac{3D_{22}}{8} - \frac{D_{11}}{2}\beta + \frac{D_{22}}{4} A(\beta)\right)v - \frac{D_{12}}{2}\beta v^2 - \frac{D_{22}}{8}\beta v^3, \tag{26}$$

$$K_2(\ln \beta) = \frac{D_{22}}{2}(1 + A(\beta)).  \qquad (27)$$

Equations (21), (26), and (27) constitute the main result of the present work. It can be seen from Eqs. (21) and (27) that $f(\beta)$ is determined by the drift, which is a mechanical quantity and can experimentally be obtained if temporal variations of the inverse temperature are measured. For example, in the case when $K_2(X_2)$ is linear in $X_2$, that is, $K_2(\ln \beta) = -[D_{22}/(2\sigma^2)](\ln \beta + c)$ with c and $\sigma$ being constants, $f(\beta)$ is found to be the log-normal distribution: $f(\beta) = \left(\beta\sqrt{2\pi\sigma^2}\right)^{-1}$ $\times \exp\left\{-\left[\ln(\beta/\beta_0)\right]^2/(2\sigma^2)\right\}$, where $\beta_0 = e^{-c}$.

$K_1(v, \ln \beta)$ in Eq. (26) is cubic in terms of $v$. This is a remarkable point in view of the fact that the Gaussian velocity distribution is obtained for the linear drift term if the background medium is fixed. Thus, the cubic nature has its origin in the dynamical treatment of the temperature.

## 3  Comment on Superstatistics

Finally, we make a comment on superstatistics in Ref. [3], which has been widely discussed in the literature.

Take a certain physical quantity, $Q = Q(V)$. Its expectation value reads

$$<Q> = \int dv \int d\beta \, Q(v) P(v, \beta).  \qquad (28)$$

Mathematically, it is possible to express it as follows:

$$<Q> = \int dv\, Q(v)\, p(v), \tag{29}$$

where $p(v)$ is given by

$$p(v) = \int d\beta\, P(v|\beta)\, f(\beta). \tag{30}$$

Superstatistics is a statistics based on $p(v)$, i.e., superposition of $P(v|\beta)$ with respect to the weight factor, $f(\beta)$. It has originally been introduced in order to obtain non-Gaussian distributions. In fact, if $f(\beta)$ is not sharply peaked at some value of $\beta$, then $p(v)$ can significantly deviate from the Maxwellian distribution. However, one should recall that the variable, $\beta$, integrated in Eq. (30) is of the slow dynamics. That is, in superstatistics, the integration over the fast variable is taken after that over the slow variable. This procedure is opposite to what the adiabatic scheme tells us. Physically, it is desirable to express $<Q>$ as follows:

$$<Q> = \int d\beta\, \tilde{Q}(\beta)\, f(\beta), \tag{31}$$

where

$$\tilde{Q}(\beta) = \int dv\, Q(v)\, P(v|\beta). \tag{32}$$

One might think that the difference between Eqs. (29) and (31) is nothing but the order of the integrations. As long as an expectation value is concerned, this is true. However, the point becomes much clearer, if the entropy is considered. To do so, we discretize both the velocity and inverse-temperature variables, so that the normalization conditions are now given as follows: $\sum_v \bar{P}(v|\beta) = 1$, $\sum_\beta \bar{f}(\beta) = 1$, where the over-bar indicates that the distribution is discrete. Substituting $\bar{P}(v,\beta) = \bar{P}(v|\beta)\bar{f}(\beta)$ into the definition of the entropy, $S[V,B] = -k_B \sum_{v,\beta} \bar{P}(v,\beta) \ln \bar{P}(v,\beta)$, we have the following celebrated relation:

$$S[V,B] = S[V|B] + S[B], \qquad (33)$$

where $S[V|B]$ and $S[B]$ are the conditional and marginal entropies given by

$$S[V|B] = \sum_\beta \bar{f}(\beta) S[V|\beta], \quad S[V|\beta) = -k_B \sum_v \bar{P}(v|\beta) \ln \bar{P}(v|\beta), \qquad (34)$$

$$S[B] = -k_B \sum_\beta \bar{f}(\beta) \ln \bar{f}(\beta), \qquad (35)$$

respectively. It is obvious from Eq. (34) that the summation over the slow variable, $\beta$, has to be performed *after* the elimination of the fast variable, *v*, in conformity with the adiabatic scheme. A relevant discussion about "conditional approach to thermodynamics" can be found in Ref. [5].

## 4   Conclusion

We have studied a kinetic theory of a nonequilibrium system governed by hierarchical dynamics organized at different time scales. To develop the discussion, we have considered, as an analytically tractable example, Brownian motion of a particle moving through a fluctuating background medium with slowly varying temperature. In our approach, not only the velocity of the particle but also the temperature has been treated dynamically. The dynamical hierarchy is realized by the adiabatic scheme analogous to the Born-Oppenheimer approximation. Developing a new method for analyzing the multivariate Fokker-Planck equation, we have derived the Maxwellian distribution modulated by the temperature fluctuations.

*Note Added*   After completion of this work, we have noticed an interesting work in Ref. [6], where separation of time scales in a Langevin equation and associated kinetic equation are discussed in connection with chemical reactions in the fluctuating environment with randomly varying temperature. This suggests a possibility of applying the present theory to such systems. Also, an anonymous referee has drawn our attention to Refs. [7,8]. The work in Ref. [7] is concerned with a stochastic volatility model, and the Fokker-Planck equation with linear drift terms is discussed. On the other hand, the authors of Ref. [8] discuss a stochastic equation only for temperature appearing in the presumed Boltzmann-Gibbs distribution. Both of Refs. [7,8] belong to studies of superstatistics, which we have carefully examined in Section 3. In fact, the volatility in Ref. [7] and the varying temperature in Ref. [8] are slow variables, but they are eliminated first in order to obtain non-exponential distributions.

**Acknowledgments**

The author would like to thank C. Beck and E.G.D. Cohen for discussions and comments. He also thanks Péter Ván for drawing his attention to the work in Ref. [9], where the Born-Oppenheimer-like approximation is discussed for a Fokker-Planck equation in the context of superstatistics. This work has been supported in part by a Grant-in-Aid for Scientific Research from the Japan Society for the Promotion of Science.

**References**

1. Risken, H.: The Fokker-Planck Equation, 2nd edn. Springer-Verlag, Berlin (1989)

2. Jacobs, K.: Stochastic Processes for Physicists. Cambridge University Press, Cambridge (2010)

3. Beck, C., Cohen, E.G.D.: Physica A **322**, 267 (2003)

4. Sakurai, J.J.: Modern Quantum Mechanics, Revised Edition. Addison-Wesley, Reading (1994)

5. Abe, S., Beck, C., Cohen, E.G.D.: Phys. Rev. E **76**, 031102 (2007)

6. Festa, C., Fonseca, T., Fronzoni, L., Grigolini, P., Papini, A.: Phys. Lett. A **117**, 57 (1986)

7. Dragulescu, A.A., Yakovenko, V.M.: Quantitative Finance **2**, 443 (2002)

8. Salazar, D.S.P., Vasconcelos, G.L.: Phys. Rev. E **86**, 050103(R) (2012)

9. Biró, T.S., Rosenfeld, R.: Physica A **387**, 1603 (2008)